# NEW EMBEDDING OF SCHWARZSCHILD GEOMETRY

# II. INTERIOR SOLUTION

Rainer Burghardt[i]

We embed the Schwarzschild interior solution in a five-dimensional flat space and show that the systems of the interior and the exterior solution are based on the same geometrical principles.

## INTRODUCTION

In a former paper [1] we have shown that it is possible to embed the Schwarzschild exterior solution in a flat five-dimensional flat space, if this method is based on a geometrical theory of a double surface. For the interior solution, it is well known that the space-like region is part of a three-dimensional sphere, but for the whole metric only *local* embeddings are worked out [2]. We will demonstrate that the time-like part of the metric may be globally embedded too, if we extend the motion of the curvature vector of Schwarzschild's parabola to the inner region. The rotation of this vector through the r-axis describes the time-surface and explains the time-like part of the interior metric.

Moreover, a system of covariant field equations is deduced from Einstein's equations, which are in close relation to the geometrical background of the theory of double surfaces.

## 1. FIVE-DIMENSIONAL FORMULATION

In [1] we have shown that a pseudo-spherical four-dimensional single surface embedded in a five-dimensional flat space can be mapped to a four-dimensional parabolic-hyperbolic double surface, representing the exterior Schwarzschild geometry. The curvature vector of the Schwarzschild parabola is connecting the parabola and its evolute (Neil's parabola).

The rotation of these two curves through the directrix (R-axis) of the parabola creates Flamm's paraboloid and the corresponding surface due to Neil's parabola. An additional rotation of these two curves through the symmetry axis (r-axis) of the Schwarzschild parabola creates the time-like double surface. The real image of these could be envisaged as parabolic-hyperbolic surface of translation, and a similar surface created by Neil's parabola. A suitable projection formalism cuts off all the geometry not necessary for describing the four-dimensional physics and reduces the five-dimensional field equations of the spherical single surface to the four-dimensional equations of what we call the physical surface.

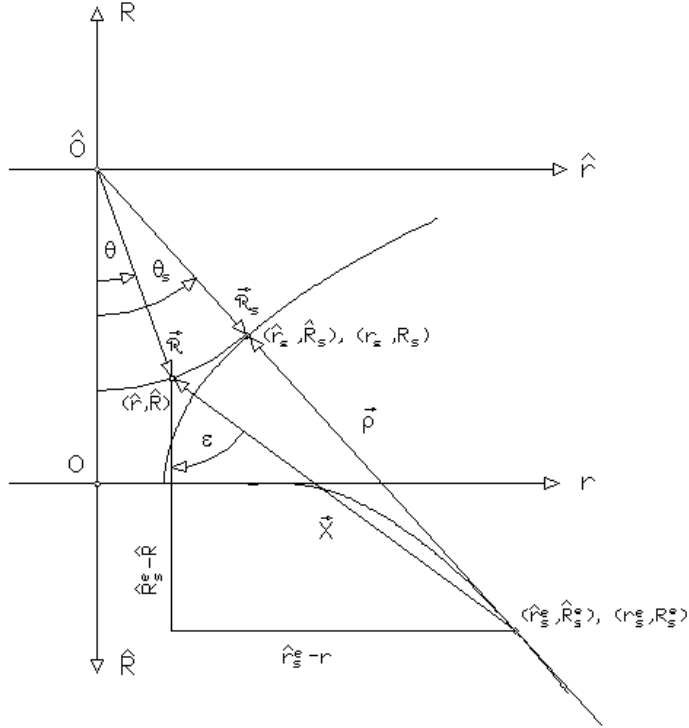

fig. 1

The curvature vector ρ of the Schwarzschild parabola plays an important role for both Schwarzschild models. The gravitational force of the exterior solution (ES) is mainly based on ρ and the motion of ρ explains most of the physics. The mass parameter M of the ES implies the existence of an interior solution (IS) that is there has to be a massive object with definite extent, which cuts off the ES at ($r_s, R_s$). $r$ and $R$ are the cartesian coordinates of the Schwarzschild parabola in the radial sections of the space-time, as can be seen in fig. 1.

For the IS we use the rectilinear co-ordinate system ($\hat{r}, \hat{R}$) and for the ES ($r, R$). From the curvature vector of the Schwarzschild parabola $\rho = \sqrt{2r^3/M}$ [1], the velocity of a freely falling object $\sin\varepsilon = -\sqrt{2M/r}$ and the relation $\varepsilon_s = -\theta_s$ we get

$$\rho_s = 2\mathcal{R}_s,  \quad (1.1)$$

$\mathcal{R}_s$ being the radius of the Schwarzschild trough. The distance from the origin of $\rho_s$ on Neil's parabola, which is the evolute of the Schwarzschild parabola, to the origin $\hat{O}$ in the IS-coordinate system is $3\mathcal{R}_s$, and the projection onto the IS coordinate lines are $3\mathcal{R}_s \cos\theta_s, 3\mathcal{R}_s \sin\theta_s$.

If we follow the motion of $\rho$, its tip moves on the Schwarzschild parabola and its origin on Neil's parabola until $\rho$ has reached the surface of a stellar object at $(r_s, R_s)$. The origin of $\rho$ now stops at $(r_s^e, R_s^e)$, while the tip of the vector moves on along the circle with radius $\mathcal{R}_s$ and origin $\hat{O}$. The motion of this vector, which we now call $X$, and its rotation through the axis $r$ and axis $R$ describes the geometry of the IS.

In a flat five-dimensional space we use two cartesian coordinate systems, $x^{i'}$ for the IS and $x^{i''}$ for the ES. If we introduce pseudo-polar co-ordinates, the components of the vector fields $X$ in both systems are

$$\begin{aligned}
X^{3''} &= X \sin\varepsilon \sin\vartheta \sin\varphi & X^{3'} &= X \sin\varepsilon \sin\vartheta \sin\varphi \\
X^{2''} &= X \sin\varepsilon \sin\vartheta \cos\varphi & X^{2'} &= X \sin\varepsilon \sin\vartheta \cos\varphi \\
X^{1''} &= X \sin\varepsilon \cos\vartheta & X^{1'} &= X \sin\varepsilon \cos\vartheta \\
X^{0''} &= X \cos\varepsilon \cos i\psi & X^{0'} &= -X \cos\varepsilon \cos i\psi \\
X^{4''} &= X \cos\varepsilon \sin i\psi & X^{4'} &= -X \cos\varepsilon \sin i\psi
\end{aligned} \qquad (1.2)$$

As $X \sin\varepsilon = \hat{r} - \hat{r}_s^e, -X \cos\varepsilon = \hat{R} - \hat{R}_s^e$ and

$$X^2 = (\hat{r} - \hat{r}_s^e)^2 + (\hat{R} - \hat{R}_s^e)^2 \qquad (1.3)$$

the vector field $X$ parametrized by (1.2) describes a family of pseudo-spheres with center $(\hat{r}_s^e, \hat{R}_s^e)$, which could be used for a single surface theory. With the help of

$$\hat{r} = \mathcal{R} \sin\theta, \hat{R} = \mathcal{R} \cos\theta, \hat{r}_s^e = 3\mathcal{R}_s \sin\theta_s, \hat{R}_s^e = 3\mathcal{R}_s \cos\theta_s \qquad (1.4)$$

we find

$$\begin{aligned}
X^{3'} &= \mathcal{R} \sin\theta \sin\vartheta \sin\varphi - 3\mathcal{R}_s \sin\theta_s \sin\vartheta \sin\varphi \\
X^{2'} &= \mathcal{R} \sin\theta \sin\vartheta \cos\varphi - 3\mathcal{R}_s \sin\theta_s \sin\vartheta \cos\varphi \\
X^{1'} &= \mathcal{R} \sin\theta \cos\vartheta - 3\mathcal{R}_s \sin\theta_s \cos\vartheta \\
X^{0'} &= \mathcal{R} \cos\theta \cos i\psi - 3\mathcal{R}_s \cos\theta_s \cos i\psi \\
X^{1'} &= \mathcal{R} \cos\theta \sin i\psi - 3\mathcal{R}_s \cos\theta_s \sin i\psi
\end{aligned} \qquad (1.5)$$

which now parametrizes a double surface. A suitable projection formalism cuts off the first three rows of the second column of (1.5). With the remaining elements the line element is

$$ds^2 = d\mathcal{R}^2 + \mathcal{R}^2 d\theta^2 + \mathcal{R}^2 \sin^2\theta d\vartheta^2 + \mathcal{R}^2 \sin^2\theta \sin^2\vartheta d\varphi^2 + [\mathcal{R}\cos\theta - 3\mathcal{R}_s \cos\theta_s]^2 di\psi^2. \qquad (1.6)$$

If we constrain the vector $\mathcal{R}$ to the Schwarzschild trough the embedding condition reads

$$\mathcal{R} = \mathcal{R}_s = const.. \qquad (1.7)$$

Introducing the co-ordinate time $dt = \rho_s d\psi = 2\mathcal{R}_s d\psi$, we get the four-dimensional line element

$$ds^2 = \mathcal{R}^2 d\theta^2 + \mathcal{R}^2 \sin^2\theta d\vartheta^2 + \mathcal{R}^2 \sin^2\theta \sin^2\vartheta d\varphi^2 - \frac{1}{4}[\cos\theta - 3\cos\theta_s]^2 dt^2 \qquad (1.8)$$

which we call the line element of the *physical surface*. Using $r = \mathcal{R}\sin\theta$, this line element also reads

$$ds^2 = \frac{1}{1-\frac{r^2}{\mathcal{R}^2}} dr^2 + r^2 d\vartheta^2 + r^2\sin^2\vartheta d\varphi^2 - \frac{1}{4}[\cos\theta - 3\cos\theta_s]^2 dt^2. \qquad (1.9)$$

Unfortunately, starting with (1.5), the calculation turns out to be rather tedious. Therefore, instead of projecting *out* superfluous terms in (1.5) we make a new ansatz for the single surface

$$\begin{aligned}
\mathcal{R}^{3'} &= \mathcal{R}\sin\theta\sin\vartheta\sin\varphi \\
\mathcal{R}^{2'} &= \mathcal{R}\sin\theta\sin\vartheta\cos\varphi \\
\mathcal{R}^{1'} &= \mathcal{R}\sin\theta\cos\vartheta \\
\mathcal{R}^{0'} &= \mathcal{R}\cos\theta\cos i\psi \\
\mathcal{R}^{4'} &= \mathcal{R}\cos\theta\sin i\psi
\end{aligned} \qquad (1.10)$$

and project *in* the last two rows of the second column of (1.5). (1.10) describes a family of pseudo-spheres with the center $\hat{O}$. The embedding condition constrains the vector $\mathcal{R}$ to the Schwarzschild through as shown in fig. 1. If we use (1.10) the metric of the flat space is

$$ds^2 = d\mathcal{R}^2 + \mathcal{R}^2 d\theta^2 + \mathcal{R}^2\sin^2\theta d\vartheta^2 + \mathcal{R}^2\sin^2\theta\sin^2\vartheta d\varphi^2 + \mathcal{R}^2\cos^2\theta di\psi^2, \qquad (1.11)$$

which also describes the metric of a family of pseudo-spheres, if $\mathcal{R}$ is kept constant. If we use (1.5), we get the metric of a family of double surfaces, parametrized by $X$

$$ds^2 = dX^2 + X^2 d\varepsilon^2 + X^2\sin^2\varepsilon d\vartheta^2 + X^2\sin^2\varepsilon\sin^2\vartheta d\varphi^2 + X^2\cos^2\varepsilon di\psi^2, \qquad (1.12)$$

the tip of $X$ being situated on one and the origin of $X$ on the other associate of the double surface. The field equations (equations for the curvature fields of the surfaces) have the same structure as the field equations for the ES [1].

## 2. THE PHYSICAL SURFACE

To find the projection formalism mentioned above, we introduce polar co-ordinate systems. The polar co-ordinates due to the vector field $\mathcal{R}$ and relation (1.10) we denote by the indices $a,b,... = \{0,1,...,4\} = \{\mathcal{R},\theta,\vartheta,\varphi,i\psi\}$, the polar co-ordinates due to $X$ and the second relation (1.2) by $\tilde{a},\tilde{b},... = \{\tilde{0},\tilde{1},...,\tilde{4}\} = \{X,\varepsilon,\vartheta,\varphi,i\psi\}$. In this co-ordinate systems the components of the vector fields $\mathcal{R}$ and $X$ are

$$\mathcal{R}^a = \{\mathcal{R},0,0,0,0\}, \ X^{\tilde{a}} = \{X,0,0,0,0\}. \tag{2.1}$$

Their differentials can be read off from (1.11, 1.12)

$$d\mathcal{R}^a = \{d\mathcal{R}, \mathcal{R}d\theta, \mathcal{R}\sin\theta d\vartheta, \mathcal{R}\sin\theta\sin\vartheta d\varphi, \mathcal{R}\cos\theta di\psi\}, \tag{2.2}$$

$$dX^{\tilde{a}} = \{dX, Xd\varepsilon, X\sin\varepsilon d\vartheta, X\sin\varepsilon\sin\vartheta d\varphi, X\cos\varepsilon di\psi\} \ . \tag{2.3}$$

These systems differ by a rotation in the $[0',1']$-plane through the angle $\omega = \varepsilon + \theta$ ($\omega$ and $\varepsilon$ having the orientation cw, but $\theta$ ccw) and an opposite orientation of their $0$-axis and $\tilde{0}$-axis respectively. By some calculations, it can be shown that

$$dX^2 + X^2 d\varepsilon^2 = d\mathcal{R}^2 + \mathcal{R}^2 d\theta^2,$$

which expresses the fact that the tip of $\mathcal{R}$ and the tip of $X$ are subjected to a common motion in the $[0',1']$-plane. The two terms in (2.2, 2.3) containing $\mathcal{R}\sin\theta$, $X\sin\varepsilon = \mathcal{R}\sin\theta - 3\mathcal{R}_s\sin\theta_s$ are easily understood by our strategy of projecting in or of projecting out the elements for the physical surface. The remaining work to be done is to project in the correct expression for the time-like part of the metric (1.11). For this purpose, we define a projector

$$\mathcal{P}_0^0 = \mathcal{P}_1^1 = \mathcal{P}_2^2 = \mathcal{P}_3^3 = 1, \ \mathcal{P} = \mathcal{P}_4^4 = \frac{\mathcal{R}\cos\theta}{-X\cos\varepsilon} = \frac{\mathcal{R}\cos\theta}{\mathcal{R}\cos\theta - 3\mathcal{R}_s\cos\theta_s}. \tag{2.4}$$

With the relations

$$dX^{\tilde{a}} = \mathcal{P}_b^{\tilde{a}} d\mathcal{R}^b, \ d\mathcal{R}^a = \mathcal{P}_{\tilde{b}}^a dX^{\tilde{b}} \tag{2.5}$$

and the embedding condition (1.7) we are able to derive the line element for the physical surface (1.8) from the line element of the single surface (1.11). Dropping the tilde, our tools for handling the interior Schwarzschild geometry are the simple relations

$$dX^a = \left(\mathcal{P}^{-1}\right)_b^a d\mathcal{R}^b, \ \partial_a = \mathcal{P}_a^b \hat{\partial}_b, \ A_{ab}^{\ c} = \mathcal{P}_a^d \hat{A}_{db}^{\ c}, \tag{2.6}$$

$dX^a, \partial_a = \partial/\partial X^a, A_{ab}^{\ c}$ being the components of the differentials of the vector field $X$ with respect to the polar system, the partial derivatives, and connexion coefficients of the embedded system and $d\mathcal{R}^a, \hat{\partial}_a = \partial/\partial \mathcal{R}^a, \hat{A}_{ab}^{\ c}$ the corresponding quantities of the single surface (1.10). We want to outline that the quantities $A_{ab}^{\ c}$ are not the connexion coefficients of a single surface but the connexion coefficients of a double surface with respect to the time-like part of the metric. ES and IS have a common intersection: Two hyperbolas of constant curvature arise by the rotation of $(r_s, R_s)$ and $(r_s^e, R_s^e)$ through the Schwarzschild standard co-ordinate line $r$ by $i\psi$. Merely the space-like part coincides with the spherical single surface. In the preceding paper, it was shown that the Riemann and Ricci tensor, as known from the Riemannian geometry, could as well describe a double surface. All we have to do is to consequently apply the relations (2.6).

The connexion coefficients of the single surface split into

$$\hat{A}_{ab}{}^{c} = \hat{M}_{ab}{}^{c} + \hat{B}_{ab}{}^{c} + \hat{C}_{ab}{}^{c} + \hat{U}_{ab}{}^{c}, \tag{2.7}$$

where

$$\hat{M}_{ab}{}^{c} = m_a \hat{M}_b m^c - m_a m_b \hat{M}^c, \quad \hat{B}_{ab}{}^{c} = b_a \hat{B}_b b^c - b_a b_b \hat{B}^c, \\ \hat{C}_{ab}{}^{c} = c_a \hat{C}_b c^c - c_a c_b \hat{C}^c, \quad \hat{U}_{ab}{}^{c} = u_a \hat{U}_b u^c - u_a u_b \hat{U}^c. \tag{2.8}$$

The vector fields have the components

$$\hat{M}_a = \left\{ \frac{1}{\mathcal{R}}, 0, 0, 0, 0 \right\}, \quad \hat{B}_a = \left\{ \frac{1}{\mathcal{R}}, \frac{1}{\mathcal{R}} \cot\theta, 0, 0, 0 \right\}, \\ \hat{C}_a = \left\{ \frac{1}{\mathcal{R}}, \frac{1}{\mathcal{R}} \cot\theta, \frac{1}{\mathcal{R} \sin\theta} \cot\vartheta, 0, 0 \right\}, \hat{U}_a = \left\{ \frac{1}{\mathcal{R}}, -\frac{1}{\mathcal{R}} \tan\theta, 0, 0, 0 \right\} \tag{2.9}$$

and $m_a, b_a, c_a, u_a$ are the unit tangent vectors on the family of the pseudo-spherical surfaces, while $n_a$ will be their normal vectors, pointing into the local 0-direction. The quantities (2.9) are the curvatures of the circular sections of the spheres (the inverse of the values of the curvature vector)

$$\left| \hat{M}_a \right| = \frac{1}{\mathcal{R}}, \left| \hat{B}_a \right| = \frac{1}{\mathcal{R} \sin\theta}, \left| \hat{C}_a \right| = \frac{1}{\mathcal{R} \sin\theta \sin\vartheta}, \left| \hat{U}_a \right| = \frac{1}{\mathcal{R} \cos\theta}. \tag{2.10}$$

Applying (2.5) to (2.7, 2.8), we get

$$M_a = \hat{M}_a, B_a = \hat{B}_a, C_a = \hat{C}_a, U_a = \mathcal{P}\hat{U}_a. \tag{2.11}$$

It is instructive to consider the geometrical background of the last relation of (2.11). At the beginning of this chapter, we have introduced two local systems $(a, \tilde{a})$, differing by a rotation through the angle $\omega = \varepsilon + \theta$. Starting with the curvatures of the X-System

$$U_{\tilde{0}} = \frac{1}{X}, \quad U_{\tilde{1}} = -\frac{1}{X} \tan\varepsilon,$$

rotating the reference system and having in mind the opposite orientation of the two 0-axis, we retrieve with

$$U_0 = -D_0^{\tilde{a}}(\omega) U_{\tilde{a}}, U_1 = D_1^{\tilde{a}}(\omega) U_{\tilde{a}}$$

the last expression (2.11). For some calculations we need the relation

$$\mathcal{P}_{|a} = (1 - \mathcal{P}) U_a. \tag{2.12}$$

## 3. THE FIELD EQUATIONS

The set of *graded* covariant[1] derivatives

---

[1] This type of covariant derivatives was discussed in [1] in greater detail

$$M_{a\|b \atop 1} = M_{a|b}, \quad B_{a\|b \atop 2} = B_{a|b} - M_{ba}{}^c B_c,$$

$$C_{a\|b \atop 3} = C_{a|b} - M_{ba}{}^c C_c - B_{ba}{}^c C_c, \tag{2.13}$$

$$U_{a\|b \atop 4} = U_{a|b} - M_{ba}{}^c U_c - B_{ba}{}^c U_c - C_{ba}{}^c U_c$$

has the advantage that a tensor of an (n-m)-dimensional subspace covariantly differentiated by the (n-m)-method of (2.8) is also a tensor of the same (n-m)-dimensional subspace. The derivatives have the nice property

$$m_{a\|b \atop 1} = 0, \; b_{a\|b \atop 2} = 0, \; c_{a\|b \atop 3} = 0, \; u_{a\|b \atop 4} = 0. \tag{2.14}$$

Inserting the relations above into the Ricci tensor

$$R_{ab} = 2\left[A_{[a \cdot b \cdot \|c]}{}^c - A_{[a \cdot b \cdot}{}^d A_{c]d}{}^c\right] \equiv 0$$

$$A_{ab}{}^c = M_{ab}{}^c + B_{ab}{}^c + C_{ab}{}^c + U_{ab}{}^c, \tag{2.15}$$

$$\Phi_{a\|b} = \Phi_{a|b} - A_{ba}{}^c \Phi_c$$

the equations for the curvature fields decouple

$$M_{b\|a \atop 1} + M_b M_a = 0, \; B_{b\|a \atop 2} + B_b B_a = 0, \; C_{b\|a \atop 3} + C_b C_a = 0, \; U_{b\|a \atop 4} + U_b U_a = 0. \tag{2.16}$$

From the five-dimensional Einstein equations $G_{ab} = R_{ab} - \frac{1}{2} g_{ab} R \equiv 0$ we get

$$M_{b\|a \atop 1} - (g_{ab} - m_a m_b) M^c{}_{\|c \atop 1} + \overset{1}{t}_{ab} = 0, \; B_{b\|a \atop 2} - (g_{ab} - b_a b_b) B^c{}_{\|c \atop 2} + \overset{2}{t}_{ab} = 0$$

$$C_{b\|a \atop 3} - (g_{ab} - c_a c_b) C^c{}_{\|c \atop 3} + \overset{3}{t}_{ab} = 0, \; U_{b\|a \atop 4} - (g_{ab} - u_a u_b) U^c{}_{\|c \atop 4} + \overset{4}{t}_{ab} = 0 \tag{2.17}$$

$$\overset{1}{t}_{ab} = M_a M_b - (g_{ab} - m_a m_b) M^c M_c, \quad \overset{1}{t}_a{}^b{}_{\|b \atop 1} = 0$$

$$\overset{2}{t}_{ab} = B_a B_b - (g_{ab} - b_a b_b) B^c B_c, \quad \overset{2}{t}_a{}^b{}_{\|b \atop 2} = 0$$

$$\overset{3}{t}_{ab} = C_a C_b - (g_{ab} - c_a c_b) C^c C_c, \quad \overset{3}{t}_a{}^b{}_{\|b \atop 3} = 0 \; .$$

$$\overset{4}{t}_{ab} = U_a U_b - (g_{ab} - u_a u_b) U^c U_c, \quad \overset{4}{t}_a{}^b{}_{\|b \atop 4} = 0$$

For these equations, the [0,4]-decomposition can easily be performed, but we apply an alternative strategy, which gives more insight into the geometrical background. $n^a$ is the unit vector pointing into the local 0-direction and

$$A_{ab} = n_{a\|b} \tag{2.18}$$

are the generalized second fundamental forms of the double surface theory. As

$$2 n_{a\|[bc]} = R_{bca}{}^d n_d \equiv 0,$$

the Codazzi equations for the double surface theory are

$$A_{a[b\|c]} = 0. \tag{2.19}$$

The generalized second fundamental form could be extracted from the connexion coefficients by

$$A_{ab}{}^c = {'A}_{ab}{}^c + A_a{}^c n_b - A_{ab} n^c, \quad A_a{}^c = A_{ab}{}^c n^b, \tag{2.20}$$

the ${'A}_{ab}{}^c$ being the four-dimensional connexion coefficients. The Ricci tensor then decomposes to

$$R_{ab} = \left[ {'R}_{ab} + 2 A_{[a}{}^c A_{c]b} \right] + 2 \left[ A_{[a}{}^c{}_{\|c]} n_b - A_{[a \cdot b\|c]} n^c \right] + 2 n_{[c} \left[ {'A}_{a]b}{}^c{}_{\|d} n^d + A_{a]}{}^d {'A}_{db}{}^c \right] \equiv 0, \tag{2.21}$$

each bracket vanishing separately. The four-dimensional covariant derivatives and the four-dimensional Ricci tensor are defined as

$$\begin{aligned}\Phi_{a\|\underline{b}} &= \Phi_{a|\underline{b}} - {'A}_{ba}{}^c \Phi_c, \quad \Phi_{a|\underline{b}} = \Phi_{a|b} - \Phi_{a|c} n^c n_b \\ {'R}_{ab} &= 2 \left[ {'A}_{[a\cdot b\cdot \|\underline{c}]}{}^c - {'A}_{[a\cdot b\cdot}{}^d {'A}_{c]d}{}^c \right]\end{aligned} \tag{2.22}$$

The last part of (3.9) stands for the change of the geometry in the local 0-direction and is less important for the physical theory. Making use of the Einstein equations, we find (n, m =1,…,4)

$${'R}_{mn} - \frac{1}{2} g_{mn} {'R} = -\kappa T_{mn}, \quad \kappa T_{mn} = 2 A_{[m}{}^s A_{s]n} - g_{mn} A_{[r}{}^s A_{s]}{}^r \tag{2.23}$$

$$A_{[m}{}^n{}_{\|n]} = 0. \tag{2.24}$$

For a decomposition of the four-dimensional Einstein tensor, we refer to paper [1]. The energy tensor of the matter is built up by the second fundamental form and is 'already geometrized'. The field equations for the matter fields $A_{mn}$ are the Codazzi equations (3.13)[2]. With the help of (1.7, 2.4, 2.9, 2.11, 3.9), we are able to calculate the matter energy tensor

$$\begin{aligned}T_{mn} &= -p \, {}^3 g_{mn} + \mu_0 u_m u_n, \quad {}^3 g_{mn} = g_{mn} - u_m u_n \\ \kappa p &= -\frac{1}{\mathcal{R}^2}(1 + 2\mathcal{P}), \quad \kappa \mu_0 = \frac{3}{\mathcal{R}^2}, \quad \mathcal{R} = \mathcal{R}_s = const.\end{aligned} \tag{2.25}$$

$p$ and $\mu_0$ being the pressure and energy density of the matter of the stellar object. The energy tensor has to satisfy the conservation law, as the left side of the field equations is free of divergence due to the Bianchi identity. It's remarkable that the conservation of energy is also a consequence of the geometrical structure of the five-dimensional theory:

---

[2] It is of some interest to include other fields in this theory by Kaluza-Klein methods. The main difference is that Kaluza-Klein theories use anholonomic rigging vectors on the surface, which lead to antisymmetric field strength, while the present theory uses holonomic rigging vectors, which lead to symmetric field strength.

$$\kappa T_{m\phantom{n}\|n}^{\phantom{m}n} = 2A_{<n}^{\phantom{<n}s}A^{n}{}_{[m\|s]>} = 0 \phantom{xx}^{3} \tag{2.26}$$

by means of the Codazzi equation. The conservation law splits into

$$p_{\|m} = -(p+\mu_0)U_m, \phantom{x} \dot{p}=0, \phantom{x} \dot{\mu}_0 = 0, \tag{2.27}$$

$U_m$ being the negative of the gravitational force.

## CONCLUSIONS

We propose a global minimal embedding of the interior Schwarzschild solution, geometrically joined to the exterior solution by use of the methods of the double surface theory. It turns out that the energy tensor of the matter has its origin in the five-dimensional structure of the geometry and is built up by the generalized second fundamental forms. Thus, the matter is already geometrized. The Codazzi equations are the field equations for the matter field.

## ACKNOWLEDGEMENT

I am indebted to Prof. H.-J. Treder for his kind interest in this work.

---

[ii] A-2061 Obritz 246, homepage: arg.at.tf, e-mail: arg@i-one.at

---

[3] "< >" denotes cyclic permutation of indices